\documentclass[%
 reprint,
 amsmath,amssymb,
 aps,pra,
]{revtex4-1}

\usepackage{graphicx}
\usepackage{dcolumn}
\usepackage{bm}

\begin{document}

\preprint{APS/123-QED}

\title{The Connection of Topology between Systems with Different Dimensions: 1D Zak Phases to 2D Chern Number, Weyl Point as the Jumping Channel for One Singularity and Nodal Line to Merge All Singularities}

\author{Qiucui Li}
\author{Xunya Jiang}
 \email{jinagxunya@fudan.edu.cn}
\affiliation{Department of Illuminating Engineering and Light Sources, School of Information Science and Engineering,
Fudan University, Shanghai 200433, China.}

\date{\today}

\begin{abstract}
The topology in different dimensions has attracted enormous interests, e.g. the Zak phase in 1D systems, the Chern number in 2D systems and the Weyl points or nodal lines in the systems with higher dimensions. It would be fantastic to find the connection of different topology in different dimensions from one simple model and reveal the deep physical picture behind them.
In this work, we propose a new model which starts from a binary-layered 1D photonic crystal, and by introducing synthetic dimensions the topology of higher dimension systems could appear.
From this model, we find that the topology of band gap and the Chern number of the 2D systems can be predicted by
the parity-switching types and the Zak phases of the 1D systems with spatial inversion symmetry(SIS), respectively. The chiral edge state is confirmed by the winding number of the reflection phase in the topological nontrivial gap.
Different types of the topological transition in higher dimensions are found,
where two bands degenerated as Weyl point or nodal line. Surprisingly, we find that the topological connection between different dimensions and  the topological
transition types in this model can be explained by the evolving of two singularities which give rise to nonzero Zak phase of the 1D systems with SIS.
When transporting one singularity between adjacent bands, the Weyl point takes the role as the instantaneous jumping channel of the singularity in the parameter
space, and then both the Zak phases of 1D systems with SIS and the Chern number of 2D systems are changed. While both singularities moves to band-gap edges from two adjacent bands, they will merge into the nodal line. The theory for such model is also constructed. We propose that such topology connection between different dimensions could be quite universal for other systems.
\begin{description}
\item[Usage]
Secondary publications and information retrieval purposes.
\item[PACS numbers]
May be entered using the \verb+\pacs{#1}+ command.
\item[Structure]
You may use the \texttt{description} environment to structure your abstract;
use the optional argument of the \verb+\item+ command to give the category of each item.
\end{description}
\end{abstract}

\pacs{Valid PACS appear here}
\maketitle

\section{\label{sec:level1}INTRODUCTION}

The topological states of different dimensions have attracted enormous interests in condensed matter physics, prominent examples including one dimensional(1D) Su-Schrieffer-Heeger model for polyacetylene, two dimensional (2D) quantum Hall states and topological insulators(TI), and three dimensional (3D) TI, topological superconductors and semi-metal with Weyl point, etc. \cite{PhysRevLett.45.494,RevModPhys.83.1057,RevModPhys.82.3045}.
In recent years, photonic crystal (PhC) has been found to be a powerful platform to easily investigate novel topological phenomena in the condensed matter physics. The SSH model, TI and Weyl points have also been found its counterpart in the photonic systems\cite{luling,PhysRevLett.100.013904,PhysRevA.78.033834}. For these topological states, symmetries act as special roles. For example, the time-reversal symmetry is very essential for 2D or 3D topological insulators, which is origin of the degeneracy at time-reversal-invariant points in Brillouin zone and the topological non-triviality of the band-gap could be represented by the connection type between these degenerated points.
Analogy to SSH model for polyacetylene\cite{RevModPhys.83.1057}, the topological structure of one-dimensional(1D) binary layered PhC characterized by Zak phases and the symmetry of eigenmodes have been studied \cite{PhysRevX.4.021017,PhysRevLett.62.2747}. More interestingly, with introduction of synthetic dimension, the topological nontrivial band-gap structure could be realized in two-dimensional(2D) or higher dimensional synthetic systems\cite{PhysRevX.7.031032,Observation,PhysRevLett.113.050402,Multi}. Analogy to Aubry-Andr¨¦-Harper model(AAH)\cite{PhysRevLett.108.220401}, 2D topological band structure with nonzero Chern number has been realized in photonic systems and the topological radiative edge state has been found in the gap where the reflection phase has nonzero winding number\cite{PhysRevLett.112.107403,PhysRevA.91.043830,PhysRevB.93.195317}.
However, for these photonic systems with synthetic dimensions\cite{PhysRevX.7.031032,PhysRevLett.112.107403,PhysRevA.91.043830,PhysRevB.93.195317}, the spatial inversion symmetry (SIS) in the parameter space is not guaranteed, so that the special status of \emph{1D systems with spatial inversion symmetry} (1D-SSIS) and its relationship with the topology of 2D or higher-dimension systems  have not been well studied. More questions, such as ``Can we judge the topology of higher-dimensional synthetic systems only by the topological properties of simple 1D-SSIS?",``What is the relationship between the Chern number of 2D synthetic systems and the Zak phases which is only well defined at 1D-SSIS?" and
``What is the mechanism behind the different topological transitions with Weyl point or nodal line for higher-dimension systems?" are waiting to answer.

\begin{figure*}[!htb]
\includegraphics[width=17cm]{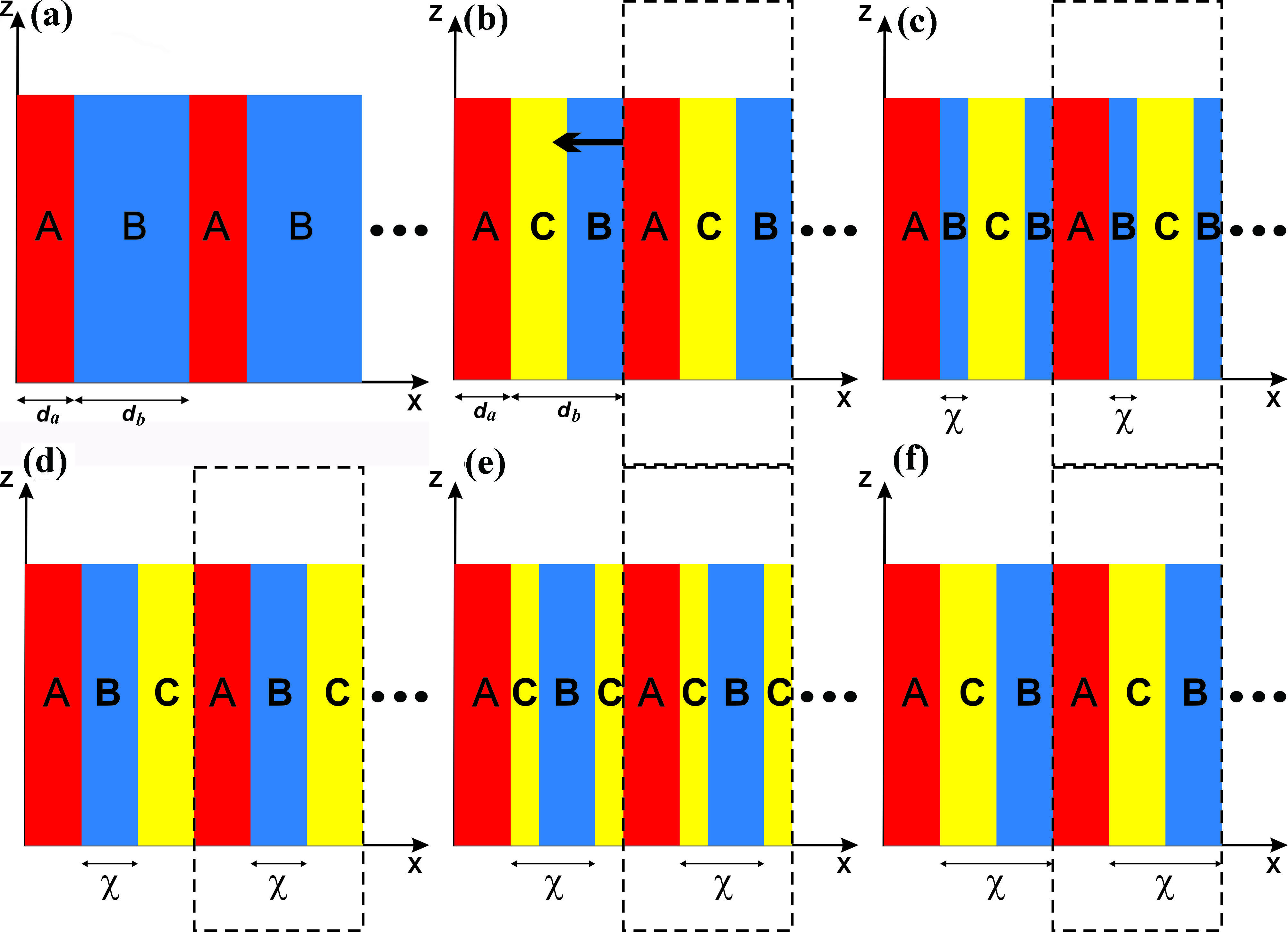}
\caption{\label{1(a)}(a) The original AB layered structure with $d_b = 2d_a$. \label{1(b)-1(f)}(b)-(f) are the five typical points during the loop with $\chi$ = $d_b/4$, $d_b/2$, $3d_b/4$ and $d_b$. (b) and (d) are the same points. The system has spatial inversion symmetry when $\chi$ = $d_b/4$ and $3d_b/4$. The width of the layer C is $d_b/2$.}
\end{figure*}

In this work, we propose a new model which starts from a simple AB binary-layered 1D photonic crystal with second gap closed, and becomes the higher dimensional one with topological nontrivial band-gap structure by introducing synthetic dimensions step by step. So the lower-dimension systems could be thought as the projection of higher-dimension systems with dimension(s) fixed at certain value(s).
At first, we suppose the original layer B of PhC cell is devided into two layers, the layer B and the layer C whose dielectric constant has a small difference from layer B, so that the second gap is lifted.  And then, the layer A is supposed to continuously move from original position through a whole PhC cell with its  displacement $\chi$ acting as the new periodic synthetic dimension.
In this model, two 1D-SSIS are guaranteed in $\chi$ dimension when layer A is at the center of layer B or layer C. Surprisingly, we find that almost all non-trivial topological properties of 2D synthetic systems are determined by the 1D systems at two 1D-SSIS. Further more, when we introduce the width and the dielectric difference of layer C as variables, two kinds of topological transition with Weyl point and nodal line are found. We reveal that all these topological phenomena of different dimensions and the connection between them are from the evolving of two singularities, which can be thought as the origin of Zak phases of 1D-SSIS. The Weyl point as the instantaneous jumping channel between bands for one singularity and the nodal line as merging of both singularities are observed.
A theory is constructed which could quantitatively describe the topology of band-gap structure for our systems with different dimensions.

 The paper is organized as following. In the first section, we introduce our model and main results from numerical simulation are presented. In the second section, we construct a theory from the effective Hamiltonian of such 2D synthetic systems based on perturbation method. In the third section, based on our effective Hamiltonian, we reveal the topology of the corresponding gap, the winding number of reflection phase and the chiral edge state. In the forth section, the relationship between the Zak phase at two 1D-SSIS and the Chern number of 2D sythetic system, and by introducing more synthetic dimensions, Weyl-point type and nodal-line-type topological transitions are found and the deep physical picture of evolving of singularities are revealed.

\section{results}
\subsection{\label{A} Model and Main Results}

\begin{figure}[b]
\includegraphics[width=8.51cm]{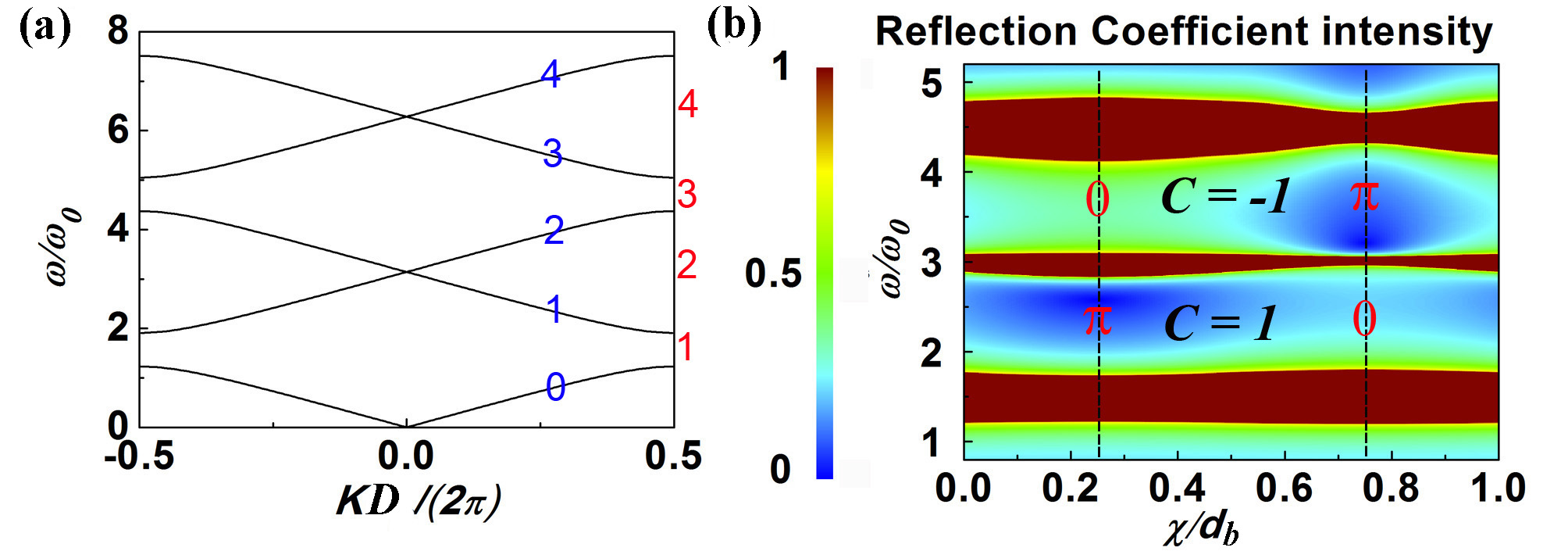}
\caption{\label{2(a)}(a) The band structure for the original AB layered structure with $\epsilon_a = 4$, $\epsilon_b=1$, $d_b = 2d_a$ and $\omega_0=c/(\sqrt{\epsilon_b}d_b)$. The bands and gaps are labeled in blue and red respectively. \label{2(b)}(b) The reflection coefficient intensity with the parameters of layer C given by $\epsilon_c=1.1\epsilon_b$ and $d_c=d_b/2$. Obviously, the degeneracy band point is lifted and there exists a gap for all the $\chi$ from $0$ to $d_b$. The Zak phases for $\chi=d_b/4$ and $\chi=3d_b/4$ are labeled in red. The two 1D-SSIS are marked by dotted lines($\chi=d_b/4$ and $\chi=3d_b/4$). The Chern number of the second and third bands are labeled in black.}
\end{figure}

We start from a dielectric binary layered structure as illustrated by Fig.~\ref{1(a)}. The width of the unit cell is denoted by $D = d_a + d_b$, where $d_a$ and $d_b$ are the width of the layer A and layer B, respectively. Its band structure is shown in Fig.~\ref{2(a)} with parameters given by $n_a=\sqrt{\epsilon_a} = 2$, $n_b=\sqrt{\epsilon_b} = 1$, and $d_b = 2d_a$. The bands and gaps are numbered respectively. Since $n_a d_a = n_b d_b$, the second gap and forth gap are closed at $K=0$ point, where \emph{K} is the Bloch wave vector.
Because the 1D system has spatial inversion symmetry, the states at the high-symmetric points of the Brillouin zone $K=0$ and $K=\pi/D$ will have certain parity, such as symmetric state (S-state) or antisymmetric state(A-state) relative to a certain inversion center. We note that, in this work, the spatial inversion center and the origin of x-coordinate is always defined as the center and the left end of layer A, respectively.

As demonstrated in \cite{PhysRevLett.62.2747,PhysRevX.4.021017}, if the upper and lower band-edge states ($K=0$ and $K=\pi/D$ ) of an isolated band have the same parity, the Zak phase of the corresponding band is $0$, otherwise it will be $\pi$ and there is a singularity in the band of 1D system.
.

\begin{figure}[!htb]
\includegraphics[width=8.51cm]{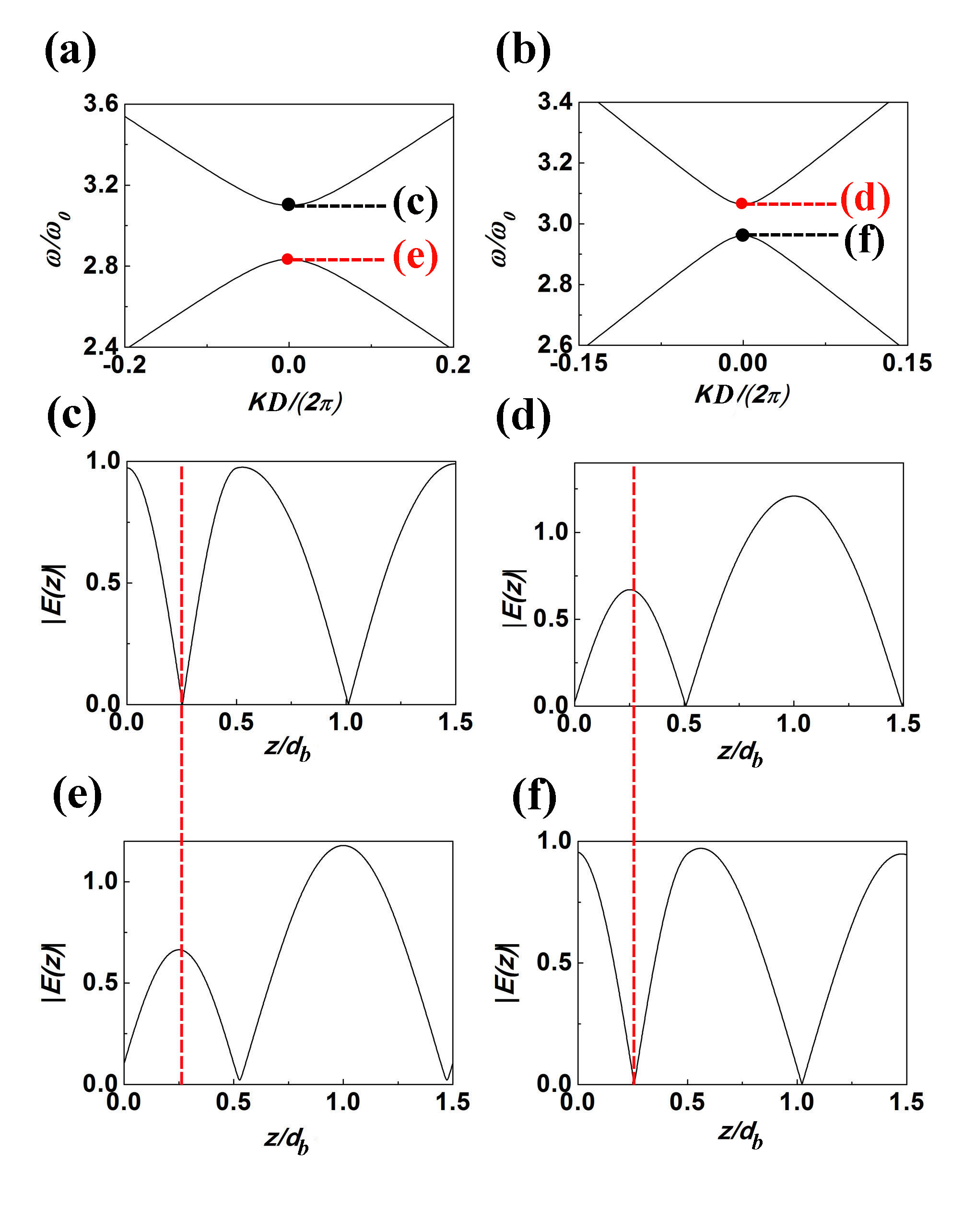}
\caption{\label{3(a)} The parameters are the same as Fig~\ref{2(b)}. (a) and (b) are the band structure for the second gap at two 1D-SSIS $\chi=d_b/4$ and $\chi=3d_b/4$ respectively. (c)-(f) are the absolute value of the electric field for the second gap edge. (c) and (d) are the upper edge states of the gap with $\chi=d_b/4$ and $\chi=3d_b/4$ respectively. (e) and (f) are the lower edge states of the gap with $\chi=d_b/4$ and $\chi=3d_b/4$ respectively. The center of the layer A is marked by red dotted line. Because the system has spacial inversion symmetry at two PWSIS, the band edge states with $K=0$ or $K=\pi$ have certain parities relative to the center of the layer A. Obviously, the parity switching type is different at the two symmetric points for this system. Symmetric state and antisymmetric state are marked by red and black dot respectively.}
\end{figure}

Next, we will propose a 2D synthetic model with a synthetic dimension by several steps. Firstly, we suppose that the layer B of our binary system is divided into two layers, layer B  and layer C. The new layer C is from $d_a$ to $d_a+d_c$, and its dielectric constant $\epsilon_c=\epsilon_b+\Delta \epsilon$ , where $\Delta \epsilon$ could (but not limited) be treated as a small perturbation. With the introduction of layer C, the closed second gap (degeneracy at K=0 point) will be lifted. Next, we continuously change the position of layer A characterised by variable $\chi$. As shown in the Fig.~\ref{1(b)-1(f)}, where
$\chi$ is the distance between the right end of layer A and the left end of layer C. When $\chi$ increases from 0 to $d_b$, the system will evolve around a closed loop. When $\chi$ is between $0$ and $d_b-d_c$, the layer A is \emph{inside} the layer B which is separated into two parts. When $\chi$ is between $d_b-d_c$ and $d_b$, the layer A is \emph{inside} the layer C which is separated into two parts. In such a way, we map a 1D system into a 2D system with the synthetic dimension $\chi$. Such model has a special advantage that two 1D-SSIS are guaranteed in $\chi$ dimension when layer A is at the center of layer B or layer C when $\chi=(d_b-d_c)/2$ or $\chi=d_b-d_c/2$, respectively.


We introduce the Chern number in our 2D synthetic system: $\int_0^{d_b}d\chi\int_{-\pi/D}^{\pi/D}dK(\partial_KA_\chi-\partial_\chi{A_K})/(2{\pi}i)$, where $A_K=\int{dxu^*(K,x)\partial_Ku(K,x)}$. $u(K,x)$ is the periodic part of magnetic field. We notice that, during the loop, two 1D-SSIS are guaranteed with $\chi=(d_b-d_c)/2$ and $\chi=d_b-d_c/2$ when the layer A is exactly at the center of layer B or at the center of layer C.
For two 1D-SSIS, 
the Zak phase can be well-defined. More important, for a certain gap of two 1D-SSIS, the lower-gap-edge state (from the lower band) and the upper-gap-edge state (from the upper band) with  $K=0$ or $K=\pi/D$ have different symmetry properties. In other words, there are two types of parity switching for the lower-gap-edge state to the upper-gap-edge state. The first type is that the lower one is S-state while the upper one is A-state which is defined as "S$\rightarrow$A-type", and the second one is the opposite case which is defined as "A$\rightarrow$S-type".

In numerical calculations, at first we set $d_c=d_b/2$ as a constant which means the width of layer C is equal as layer B and the two 1D-SSIS are the cases with $\chi=d_b/4$ and $\chi=3d_b/4$. Based on the standard transfer-matrix method, we can obtain the synthetic 2D band-gap structure in $K-\chi$ phase space, as shown in Fig.~\ref{2(b)} where the Zak phase at the two 1D-SSIS with($\chi=(d_b/4$ and $\chi=3d_b/4$) and the Chern number of bands  are signed.

After numerical simulations and theoretical derivations, some universal relations between topology of synthetic systems with different dimensions are found, which is the most important results of this work.


First, for a certain gap, we find that, if the parity switching types of gap-edge states at the two 1D-SSIS $\chi=(d_b-d_c)/2$ and $\chi=d_b-d_c/2$ are different, the 2D gap will be topologically nontrivial and contributes -1 and 1 (or reverse) to the Chern numbers of the upper and lower bands respectively. As shown in Fig.~\ref{3(a)} for second gap, at $\chi=(d_b-d_c)/2$ the parity switching from lower to upper gap edge is S$\rightarrow$A-type while at $\chi=d_b-d_c/2$ the parity switching is A$\rightarrow$ S-type. On the contrary, for the first gap and the third gap, the same parity-switching type appears at the two 1D-SSIS (both are S$\rightarrow$A-type or A$\rightarrow$S-type), so that the gap will contribute no Chern number to bands.
A strict mathematic proving of topologically non-triviality of 2D gap will be given based on the perturbation theory.

Second, the different types of parity-switching will also lead to the nonzero winding number of $\bm{\beta}$ which will be defined later for the corresponding gap, which could also be revealed by the winding number of reflection phase and is the sufficient and necessary condition for the existence of the chiral edge states.

Third, for both upper and lower bands adjacent to the lifted gap, the Chern number of the bands can be obtained by the Zak phases of 1D systems at two 1D-SSIS ($\chi=(d_b-d_c)/2$ and $\chi=d_b-d_c/2$) by:
\begin{equation}
\label{Eq:1}
C={\rm sgn}[\epsilon_c-\epsilon_b] \times (Z(\chi=(d_b-d_c)/2)-Z(\chi=d_b-d_c/2))/\pi
\end{equation}
The physical reason behind this relation is that both Zak phase of 1D systems and the Chern number of 2D synthetic system are from two singularities of the field, whose amplitude is zero in transfer-matrix gauge \cite{PhysRevX.4.021017}, in the upper band or the lower band at $\chi=(d_b-d_c)/2$ or $\chi=d_b-d_c/2$, respectively.

Forth, if we introduce the width $d_c$ and the dielectric difference $\Delta_\epsilon = \epsilon_c -\epsilon_b $ of layer C  as variables, we find two kinds of topological transition of band-gap structure in this model. One kind is that two bands touch each other as a Weyl point at one 1D-SSIS and while the other kind is that two bands connect by a ``nodal line" (or nodal surface) in higher-dimension parameter space. The physics behind all these topological phenomena is from the evolving of two singularities at 1D-SSIS which can be thought as the origin of Zak phases. At the Weyl-point-type  transition, the Weyl point acts as the instantaneous jumping channel for one singularity from one band to the other one, and the Zak phase 1D-SSIS and the Chern number of synthetic 2D system are changed. At the nodal-line-type transition, both singularities move to the band edges and merge into the nodal line.

\subsection{\label{B}Hamiltonian Based on The Perturbation Method}

\begin{figure}[!htb]
\includegraphics[width=8.51cm]{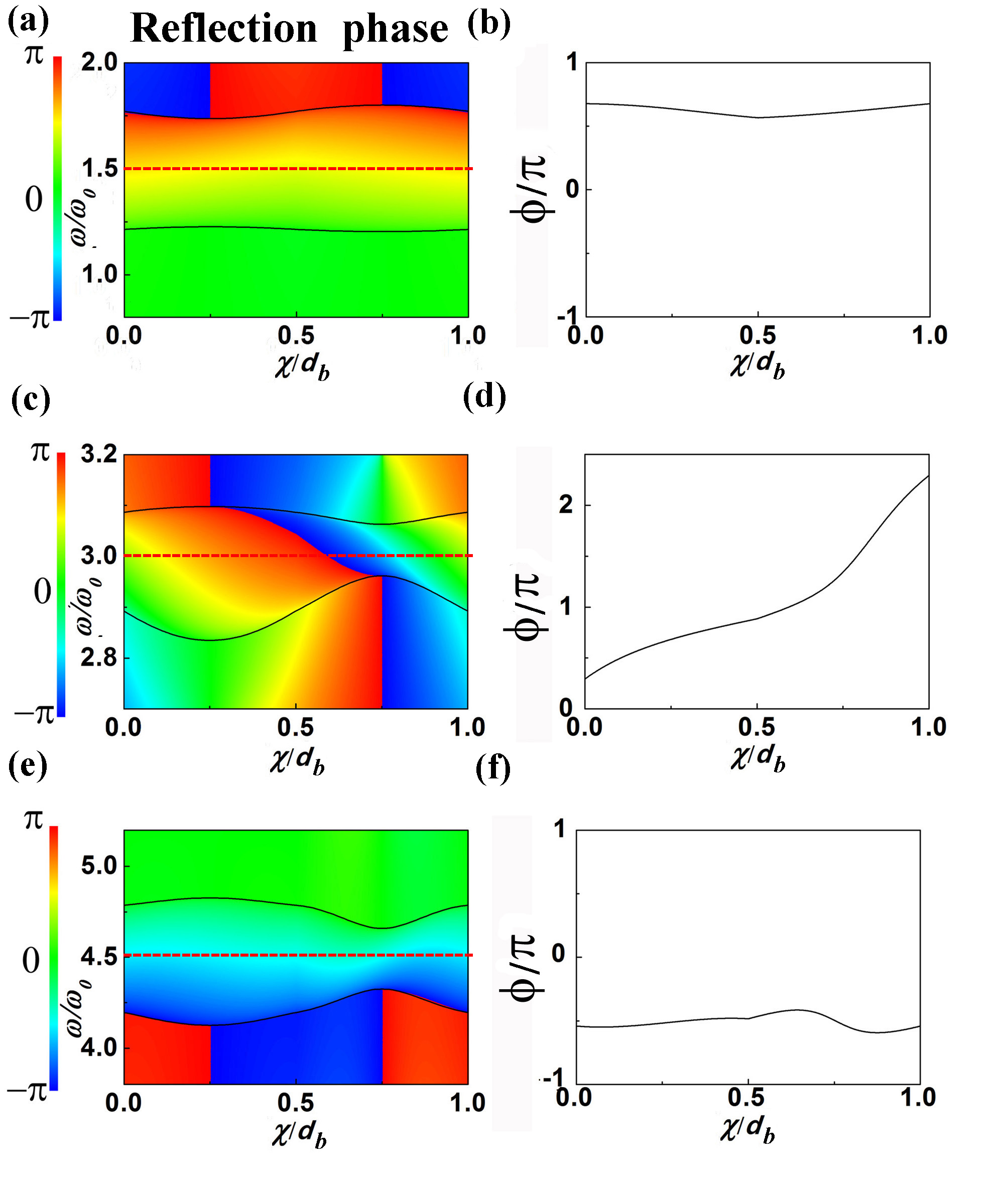}
\caption{\label{fig_4} The parameters are the same as Fig~\ref{2(b)}. (a), (c) and (e) are the reflection phase for the first, second and third gap. (b), (d) and (f) are the reflection phase for $\omega/\omega_0$ = 1.5, 3 and 4.5. The winding number of the reflection phase are 0 for the first and third gap and 1 for the second gap.}
\end{figure}

In this section, we construct a theory from the effective Hamiltonian of such 2D synthetic systems based on perturbation method. The effective $2\times2$ Hamiltonian will be obtained by the bases of the magnetic field $H(x)$ with different parity relative to layer A. From such a theory, we can clearly show the topological characters of band-gap structures of 2D synthetic systems.

To derive the effective Hamiltonian, we start from the original 1D AB-layered system. The magnetic field $H(x)$ in y-direction at the frequency of $\omega$ is determined by the Maxwell equation:
\begin{equation}
\nabla\times\frac{1}{\epsilon(x)}\nabla\times H(x)=\frac{\omega^2}{c^2}H(x)
\end{equation}

Since $n_ad_a = n_bd_b$, the second gap is closed at $K=0$ point. We have $k_b=\pi/d_b$ and $k_a=\pi/d_a$.
At $\emph{K}=0$ point, $H(x)$ has certain parity relative to the center of layer A. As we mentioned, the symmetric state is called as "S-state" while the antisymmetric state as "A-state". From Maxwell equation, we obtain that the normalized S-state has the form:

\begin{subequations}
\label{eq:whole1}
\begin{equation}
-N_1k_a\cos{k_a(x-\frac{d_a}{2})},(0 \leq x \leq d_a) \label{subeq:1}
\end{equation}
\begin{equation}
N_1k_b\cos{k_a(x-d_a-\frac{d_b}{2})},(d_a \leq x \leq d_b+d_a) \label{subeq:2}
\end{equation}
\end{subequations}

Eq.~\ref{subeq:1} and ~\ref{subeq:2} are the magnetic field(\emph{H} filed) within layer A and layer B respectively.

The antisymmetric state denoted by A-state is given by
\begin{subequations}
\label{eq:whole2}
\begin{equation}
-N_2k_b\\sin{k_a(x-\frac{d_a}{2})},(0 \leq x \leq d_a) \label{subeq:3}
\end{equation}
\begin{equation}
N_2k_b\\sin{k_b(x-d_a-\frac{d_b}{2})},(d_a \leq x \leq d_b+d_a) \label{subeq:4}
\end{equation}
\end{subequations}
where $k_i = n_i\omega/c$, $n_i=\sqrt{\epsilon_i\mu_i}$, $N_1=1/\sqrt{k_a^2\frac{d_a}{2}+k_b^2\frac{d_b}{2}}$ and $N_2=1/\sqrt{k_b^2\frac{d_a}{2}+k_b^2\frac{d_b}{2}}$ are the normalized coefficients.

These states given by Eq.~\ref{eq:whole1} and~\ref{eq:whole2} will be used as the basis in our derivation of effective Hamiltonian. For small deviation $K$ around $K=0$, these bases should be added a phase factor $exp{(iKx)}$.

Next, as we did in constructing our 2D synthetic system in the second paragraph, we introduce the perturbation layer C. We treat the deviation of $\epsilon_c$ from $\epsilon_b$ as a perturbation$\Delta\epsilon$, so we have $1/\epsilon_c=1/\epsilon_b-\Delta\epsilon/\epsilon_b^2$. Based on formal $\bm{K\cdot P}$ method,
the elements of effective Hamiltonian can be obtained by
$ \langle{H(K,x)|\frac{d}{dx}\frac{1}{\epsilon(x)}\frac{d}{dx}|H(K,x)}\rangle $.

 After some derivation, we obtain the effective Hamiltonian as:
 \begin{equation}
 \label{eq:5}
\left(
  \begin{array}{cc}
    k_b^2+t(p(\chi)-2k_bd_c)N_1^2 & \frac{-2ik_bK}{\sqrt{c_1}}+tq(\chi)N_1N_2\\
    \frac{2ik_bK}{\sqrt{c_1}}+tq(\chi)N_1N_2 & k_b^2+t(-p(\chi)-2k_bd_c)N_2^2\\
  \end{array}
\right)
\end{equation}
where
\[c_1=[(n_ad_a)^2+(n_bd_b)^2+(\frac{\sqrt{\epsilon_a}}{\sqrt{\epsilon_b}}+\frac{\sqrt{\epsilon_b}}{\sqrt{\epsilon_a}})n_ad_an_bd_b]/D^2\],
t=$k_b^3\Delta\epsilon/4\epsilon_b^2$, and
\begin{subequations}
\begin{equation}
p(\chi)=\sin2k_b\chi-\sin2k_b(\chi-d_b+d_c)\label{subeq:4}
\end{equation}
\begin{equation}
 q(\chi)=\cos2k_b(\chi-d_b+d_c)-\cos2k_b\chi\label{subeq:4}
\end{equation}
\end{subequations}
The Hamiltonian can be expressed as the Dirac form:
\begin{equation}
H_{eff}=\alpha\sigma_0+\bm{\beta\cdot\sigma}
\end{equation}
where $\sigma_0$ is the identity matrix and $\bm{\sigma}$ is the Pauli matrix, while $\alpha$ and the vector $\bm{\beta_{K,\chi}}$ are defined as the coefficients of them:
\begin{equation}
\alpha=k_b^2+\frac{t(p(\chi)-2k_bd_c)N_1^2}{2}+\frac{t(-p(\chi)-2k_bd_c)N_2^2}{2}
\end{equation}
\begin{eqnarray}
\label{eq:9}
\bm{\beta}_{K,\chi}=[tq(\chi)N_1N_2,\frac{2k_bK}{\sqrt{c_1}},&&\frac{t(p(\chi)-2k_bd_c)N_1^2}{2}\nonumber \\
&&-\frac{t(-p(\chi)-2k_bd_c)N_2^2}{2}]
\end{eqnarray}

\subsection{\label{C}Symmetry, Chern Number and Reflection Phase}
\begin{figure}[!htb]
\includegraphics[width=8.51cm]{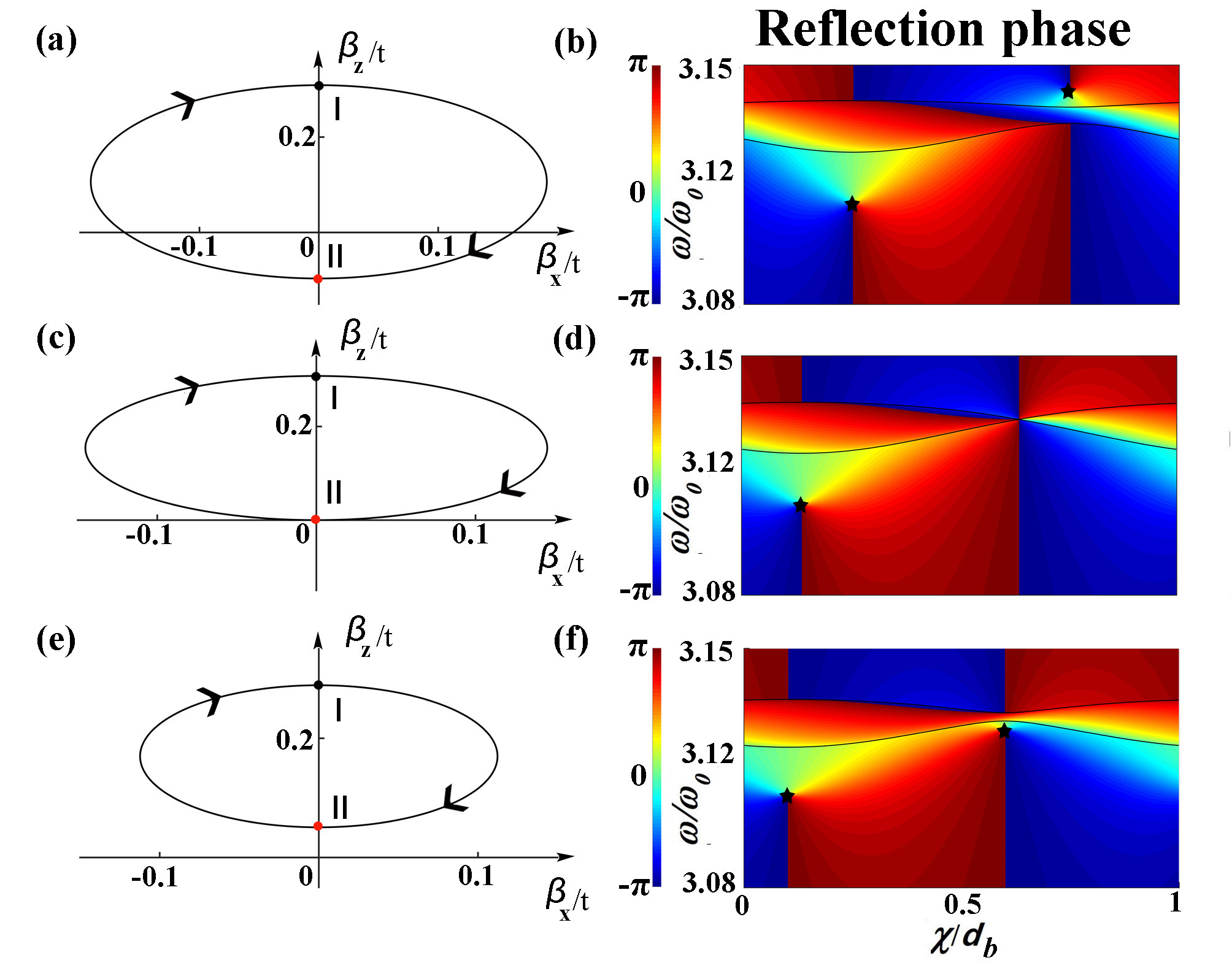}
\caption{\label{fig_5}The parameters are given by: $d_c=d_b/2$, $\epsilon_c=1.01\epsilon_b$ for (a),(b), $d_c=0.7254d_b$, $\epsilon_c=1.01\epsilon_b$ for (c),(d) and $d_c=0.8d_b$, $\epsilon_c=1.01\epsilon_b$ for (e),(f). (a) The trajectory of $\bm{\beta_{K=0,\chi}}$ with $\chi$ increasing from 0 to $d_b$. It rotates clockwise. $\bm{\beta_{K=0,\chi=d/4}}$ and $\bm{\beta_{K=0,\chi=3d/4}}$ are marked by black and red dot respectively. (b) Reflection phase of the second gap. Solid line represents the gap edge.}
\end{figure}

Then, based on the effective Hamiltonian and contour integral, we can clearly show the topological structure of 2D our synthetic system.  First, at the central frequency of the second gap with $k_b=\pi/d_b$ and $k_a=\pi/d_a$, in $\{\chi, K \}$ space we suppose there is a periodically evolving path which $\chi$ changes from $0$ to $d_b$ with $K=0$. Along this path, $K=0$ results $\beta_y=0$, so that the vector $\bm{\beta_{K=0,\chi}}$  transports around a loop in the $\beta_x$-$\beta_z$ plane. As we demonstrated previously, when  $d_c=d_b/2$, the second gap is topologically nontrivial. But if we treat $d_c$ as a variable, there could be three typical kinds of evolving loops according to the effective Hamiltonian as shown in Fig.~\ref{fig_5}. The two 1DSSIS  $\chi=(d_b-d_c)/2$ and $\chi=d_b-d_c/2$ correspond to two intersection points of the loop and the axis $\beta_z$, which is denoted by point I and point II respectively as shown in Fig.~\ref{fig_5}(a) Fig.~\ref{fig_5}(c) and Fig.~\ref{fig_5}(e). The topological property is characterized by the fact that if the loop encloses the origin or not. It is obvious that if the clockwise (or anticlockwise) loop encloses the origin $\bm{\beta}=0$, two vectors $\beta_I$ and $\beta_{II}$ are in the opposite direction, so that they must have different sign and the winding number of $\bm{\beta_{K=0,\chi}}$ of the loop will be -1 (or 1). For the case of Fig.~\ref{fig_5}e, the loop doesn't encloses the origin, so that two vectors $\beta_I$ and $\beta_{II}$ are in the same direction and the winding number of the loop is zero. From the effective Hamiltonian, it could be derived that if $d_c$ satisfies a special condition $(N_1^2-N_2^2)k_bd_c+(N_1^2+N_2^2)\sin{k_bd_c}=0$, the 2D synthetic system is on a critical condition with $\beta_{II}=0$ according to Eq.~\ref{eq:9}. Now the loop of $\bm{\beta_{K=0,\chi}}$ will be tangent to the axis $\beta_x$, as shown in Fig.~\ref{fig_5}(c), which corresponds to the transmission of band-gap structure from topological trivial case to nontrivial one. The second gap will be closed at
$\chi=d_b-d_c/2$ as shown in fig.~\ref{fig_5}d, which is a Weyl point actually and will be discussed later.

It is more interesting that we can obtain the wave-functions and the Chern number of upper and lower bands from the effective Hamiltonian too, due to its simple form.
Using our effective Hamiltonian, valid for small $K$ and $\Delta\epsilon$, we calculate the gap's contribution to the Chern number of the upper and lower bands. When the gap is topological nontrivial and the winding number of $\bm{\beta_{K=0,\chi}}$ is $+1$, the gap will contribute $-1$ and $+1$ to the Chern number of the upper and lower bands, respectively. For the convenience of discussion, we define it as the "gap winding number".
For the topological nontrivial case of Fig.~\ref{fig_5}a, according to the effective Hamiltonian we can easily find that the parity-switching types (S$\rightarrow$A or A$\rightarrow$S) of gap-edge wave-functions at points $\beta_I$ and $\beta_{II}$ (corresponding to $\chi=(d_b-d_c)/2$ and $\chi=d_b-d_c/2$) are different. Since $\beta_x=\beta_y=0$, the $H_eff$ becomes $\beta_z \sigma_z + \alpha\sigma_0$, and the gap-edge state of upper band is S-state $(1,0)^T$ while the gap-edge state of lower band is A-state $(0,1)^T$ for positive $\beta_z$. The reverse applies for the negative $\beta_z$.
 Because the loop in the Fig.~\ref{fig_5}a encloses the origin, the $\beta_I$ and $\beta_{II}$ must have different sign and the parity-switching type must be different at $\chi=(d_b-d_c)/2$ and $\chi=d_b-d_c/2$. For the topological trivial case shown in Fig.~\ref{fig_5}c, since $\beta_I$ and $\beta_{II}$ must have the same sign, the parity-switching types are the same at $\chi=(d_b-d_c)/2$ and $\chi=d_b-d_c/2$.
In summary, the first general conclusion in section \ref{A} can also be strictly proved from our effective Hamiltonian.

As stated in\cite{PhysRevA.91.043830}, the topological properties of gap are discussed based on the winding number of reflection phase $\phi$ for the frequency inside gap. Next we will illustrate that the winding number of $\bm{\beta}$ is exactly same as the winding number of reflection phase $\phi$. From the effective Hamiltonian and the evolving path at gap central frequency with $\chi$ from $0$ to $d_b$, we can make analogy to an electronic spin state led by the external magnetic field adiabatically in a closed evolving path. From the derivation of Hamiltonian, we can easily see that the wave-function bases of even and odd state correspond to the spin-up $(1,0)^T$ and spin-down state $(0,1)^T$ respectively, and ``the external magnetic field" $\bm{\beta}$  rotates in the $\beta_x$-$\beta_z$ plane. On the other hand, the reflection coefficient inside gap is given by $r_{\chi}=e^{i \phi_{\chi}}$, where $\phi_\chi$ is the reflection phase.
Correspondingly, the \emph{H} field in the layer A has the form: \begin{equation}
H(K=0,\chi)=\frac{\sqrt{2}}{2}e^{i{k_a(x-da/2)}}+\frac{\sqrt{2}}{2}e^{i\phi_{\chi}}e^{-i{k_a(x-da/2)}}
\end{equation}
which could be expressed in even and odd basis as $\frac{1}{2}(1+\exp{i\phi_\chi})\cos(k_a(x-d_a/2))+\frac{1}{2}(i-i\exp{i\phi_\chi})\sin(k_a(x-d_a/2))$. Then the expectation for $S_z$ and $S_x$ is $\langle{H(K=0,\chi)}|\sigma_z|{H(K=0,\chi)}\rangle=\cos{\phi_\chi}$ and  $\langle{H(K=0,\chi)}|\sigma_x|{H(K=0,\chi)}\rangle=\sin{\phi_\chi}$. So the reflection phase $\phi_\chi$ is nothing but the angle between $\langle\textbf{S}\rangle$ and $\beta_z$ axis in our model. In the evolving process, ``the external magnetic field" drags the ``spin" following it to rotate in a closed path. Hence, the winding number of reflection phase could be reinterpreted in geometric picture from our theory. Based on the discussion aforementioned, the sufficient and necessary condition to achieve nonzero winding number of reflection phase is to introduce different band parity-switching types at two 1D-SSIS to the corresponding gap. 

Analogue to the quantum Hall systems, the``bulk-edge correspondence" theory still works in our 2D synthetic systems that if the bulk system is topological nontrivial with a nonzero Chern number, then chiral edge states appear in the gap characterized by its unidirectional propagation. There are chiral edge states appearing in the nontrivial gap of our system. To simplify the discussion, we suppose that our 1D PhC is attached to a perfect metal slab and the chiral edge states should appear at the interface between them. The condition for the chiral edge states is:

\begin{equation}
\label{eq:11}
\phi_{PhC}+\phi_R=2m\pi,   m \in \mathbb{Z}
\end{equation}
where reflection phases of the our PhC and reflecting metal are denoted by $\phi_{PhC}$ and $\phi_R$, respectively. Because the $\phi_R$ of perfect metal is $\pi$, the phase cut line depicted as $\pi$ reflecting phase in Fig.~\ref{fig_4}c, which traverses the gap, represents the properties (frequency and the $\chi$) of edge states at such interface. For the first and third gap such edge state is absent. From the slope of the phase cut line, we can also see that the edge state is ``unidirectional" as quantum Hall effect. The "time-reversal-symmetry-broken" in our 2D synthetic system is from the fact that the layer A is moved in one direction, which is different from moving in the other direction, so that transformation "$\chi \rightarrow -\chi$ is not symmetric.
Actually, no matter what reflector the PhC is attached to, e.g. another PhC or a meta-surface, the Eq.~\ref{eq:11} can always be satisfied at a certain $\chi$ for any frequency inside the gap, because the reflecting phase for the frequency in the topological nontrivial gap covers the whole $[0, 2\pi)$ when $\chi$ passes a cycle(with nonzero winding number) as shown in Fig.~\ref{fig_4}.


\subsection{\label{D}Chern Number, Zak Phase and Topological Transitions with Weyl Point and Nodal Line}

In this section, first we will study the subtle relationship between the Zak phases of 1D-SSIS at $\chi=(d_b-d_c)/2$ and $\chi=d_b-d_c/2$ and the Chern number of the 2D synthetic system,shown in Eq.~\ref{eq:5}. The relation is based on the fact that both Zak phases and Chern number are results of the same singularities in the Brillouin zone. Further more, by tuning the parameters  $d_c$ and $\Delta \epsilon$, the topological transitions of the band-gap structure of synthetic systems with different dimensions is observed. Surprisingly, we reveal that, at topological transition, different types of topological transition are from the evolving of two singularities at 1D-SSIS.


As shown in \cite{PhysRevLett.62.2747}, the Zak phase of bands is only well defined at 1D periodic systems with SIS, corresponding to two 1D-SSIS with $\chi=(d_b-d_c)/2$ and $\chi=d_b-d_c/2$ in our model. The non-zero Zak phase of such 1D systems is caused by the singularity, at which the coefficients of forward and backward wave equal to zero simultaneously and $u_{K}$ experiences $\pi$ phase jump with the gauge defined by standard transfer-matrices in\cite{PhysRevX.4.021017}, where $u_{K}$ is the periodic part of the H-field and $H_K(x) = u_K(x)exp(iKx)$. Based on the same gauge, the singularity manifests itself as the phase vortex point in our 2D synthetic system as shown in Fig.~\ref{fig_5}. Such singularities give rise to not only non-zero Zak phase of 1D-SSIS at $\chi=(d_b-d_c)/2$ and $\chi=d_b-d_c/2$, but also non-zero Chern number in our 2D synthetic system. With the gauge of standard transfer-matrices in \cite{PhysRevX.4.021017}, Chern number of a certain band equals to the sum of the contour integral $\oint(d\chi{A_\chi}+dK{A_K})/(2{\pi}i)$ around all the singularities in this band. We find that the result of the contour integral around the phase vortex point equals to the minus winding number of the reflection phase around it. In such a way, the relationship of Eq.~\ref{Eq:1} between the Chern number of 2D synthetic system and the Zak phases of 1D-SSIS could be constructed. 

Next, we introduce the width $d_c$ and the deviation of dielectric constant $\Delta \epsilon$ of layer C as new variables and more complex topological phase transitions are found. When we increase the width of layer C  $d_c$ from $d_b/2$ to a larger value, one singularity gradually moves from the lower band to the upper band as depicted in Fig.~\ref{fig_5}.
At the critical case of topological transition with $d_c$ is $d_c=0.725 d_b$, the upper band and the lower band become degenerated at $\chi=d_b-d_c/2$.
Now we illustrate this degenerated point is a Weyl point, which could be characterized by a standard Hamiltonian in a generalized parameter space spanned by $\{\chi,K,d_c\}$. Exactly as two bands touching each other, one singularity just arrives at this degenerated point which is the only bridging channel between two bands. After that critical case, if we increase $d_c$ further, two bands separate again with a gap between them and two singularities are "locked" in one band. In other words, the Weyl point is the instantaneous jumping channel for one singularity between two bands.
From our effective Hamiltonian given by Eq.~\ref{eq:5}, we can derive that the degenerated point locates at $\chi=d_b-d_c/2$ (or $\chi=(d_b-d_c)/2$ ), $K=0$, and $d_c = d_0$ which satisfies $(N_1^2-N_2^2)k_bd_0+(N_1^2+N_2^2)\sin{k_bd_0}=0$ and for our system it is at $0.725 d_b$. If we define $\delta d=d_c-d_0$ and $\delta\chi=\chi-(d_b-d_c/2)$, the effective Hamiltonian can be expanded with respect to $(\delta\chi,K,\delta d)$ around the degenerated point and has the form:
\begin{equation}
\label{eq:12}
H_w=(C+D\delta{d})\sigma_0+{\delta\chi}v_{\chi}\sigma_x+Kv_{K}\sigma_y+\delta{d}{v_d}\sigma_z
\end{equation}
where $C=k_b^2-t[(N_1^2-N_2^2)\sin{k_bd_0}+(N_1^2+N_2^2)k_bd_0]$, $D=-k_bt[(N_1^2-N_2^2)\cos{k_bd_0}+(N_1^2+N_2^2)]$, $v_{\chi}=-4N_1N_2k_bt\sin{k_bd_0}$, $v_{K}=2k_b/\sqrt{c_1}$ and $v_d=k_bt[N_2^2-N_1^2-\cos{k_bd_0}(N_1^2+N_2^2)]$. Obviously, we have illustrated that the degenerated point of two bands is a Weyl point.
As depicted in Fig.~\ref{fig_5}, when $\delta{d}$ is tuned from positive to negative, one singularity with reflection phase winding number $2\pi$ moves from lower band to upper band, hence the Zak phase of 1D-SSIS and the  Chern number 2D synthetic system of the lower band changes from $0$ to $1$. The Weyl point shown in Fig.~\ref{fig_5} has a positive charge by the standard definition.

At last, with introducing the variable $\Delta \epsilon$, a different type of topological transition, which is signed by a "nodal line" (or "nodal surface" ) similar as the case of electronic systems \cite{PhysRevB.92.081201}, could be revealed in our model from a higher-dimension parameter space $\{ \chi, K, d_c, \Delta\epsilon \}$. Before the discussing of general condition for such topological transition, it is instructive to go back to the start point of our model, where the system is a simple binary AB layered 1D system.  From the view of parameter space $\{ \chi, K, d_c, \Delta\epsilon \}$, the binary AB layered system could be view as  $\Delta\epsilon=\epsilon_c-\epsilon_b=0$ (or $d_c=0$), and the second gap is always closed for all values of $\{ \chi, K, d_c \}$. Such closed gap at certain frequency forms a nodal line in $\{ \chi, K \}$ space or "nodal surface" in higher dimension space $\{ \chi, K, d_c, \Delta\epsilon \}$. As we have demonstrated, by introducing a new layer C, the nodal line is opened as a topologically nontrivial gap, so that it is a topological transition too. Obviously, the nodal surface with $\Delta\epsilon=\epsilon_c-\epsilon_b=0$ is a very special and seemingly trivial case, since the nodal surface is a "flat one" which is independent of the dimensions $\{ \chi, K, d_c \}$.  Actually, the most general condition of nodal surfaces in our model is that the ratio of optical path of layers satisfies
\begin{equation}
k_a d_a:k_b d_b:k_c d_c=l:m:n
\end{equation}
where $l, m, n \in \mathbb{N^+}$. Then, the $l+m+n$th gap will turn into a nodal line at the frequency $w_{l+m+n}=(l+m+n)\pi{c}/(n_ad_a+n_bd_b+n_cd_c)$. We emphasize that, since now $d_a$, $d_b$ and $d_c$ are none-zero number and the materials of layers are different too, so that the nodal surface is a curved one in dimensions of $\{ K, d_c, \Delta\epsilon \}$ except the dimension $\{ \chi \}$, since the gap is always closed at certain frequency when the layer A is moved through a whole cell.

For the nodal-line-type topological transition, what will happen for these singularities of our model? From the numerical simulation, we find that, if we tune the parameter(s) of our system to the critical condition of nodal surface, both singularities inside the upper and lower bands at two 1D-SSIS will move to the gap-edge gradually and merge into the nodal line, and then they will emerge again when the parameter(s) passes the critical value. Physically, it is natural to imagine a general picture of the nodal-line type topological transition, in which the singularities from different cases with certain symmetry would merge into a line at certain frequency, and then the gap becomes totally closed and the topological non-triviality becomes extinct.

It's very introductive if we retrospect this work from the parameter space $\{\chi, K, d_c, \Delta\epsilon \}$ step by step. At first, we introduce A-B binary layered system
and the second gap is closed as a nodal line in whole parameter space. The layer B is separated into two layers, one still with original $\epsilon_b$ while the other (layer C) with $\epsilon_c=\epsilon_b + \Delta\epsilon $ and second gap is lifted. Then, the parameter $\chi$ describing the shift of layer A is brought in, so that a 2D synthetic system with topologically nontrivial gap is constructed. We find that all topological properties of 2D band-gap structure can be predicted by two 1D-SSIS with $\chi=(d_b-d_c)/2$ and $\chi=d_b-d_c/2$. Further more, when we scan the parameter space $\{\chi, K, d_c, \Delta\epsilon \}$ , two types of topological transition appears.  For Weyl-point-type transition, one singularity of 1D-SSIS jump from one band to the other band through the Weyl point, while for the nodal-line-type transition, both singularities of 1D-SSIS merge into the nodal line together.



\begin{figure}[!htb]
\includegraphics[width=8.51cm]{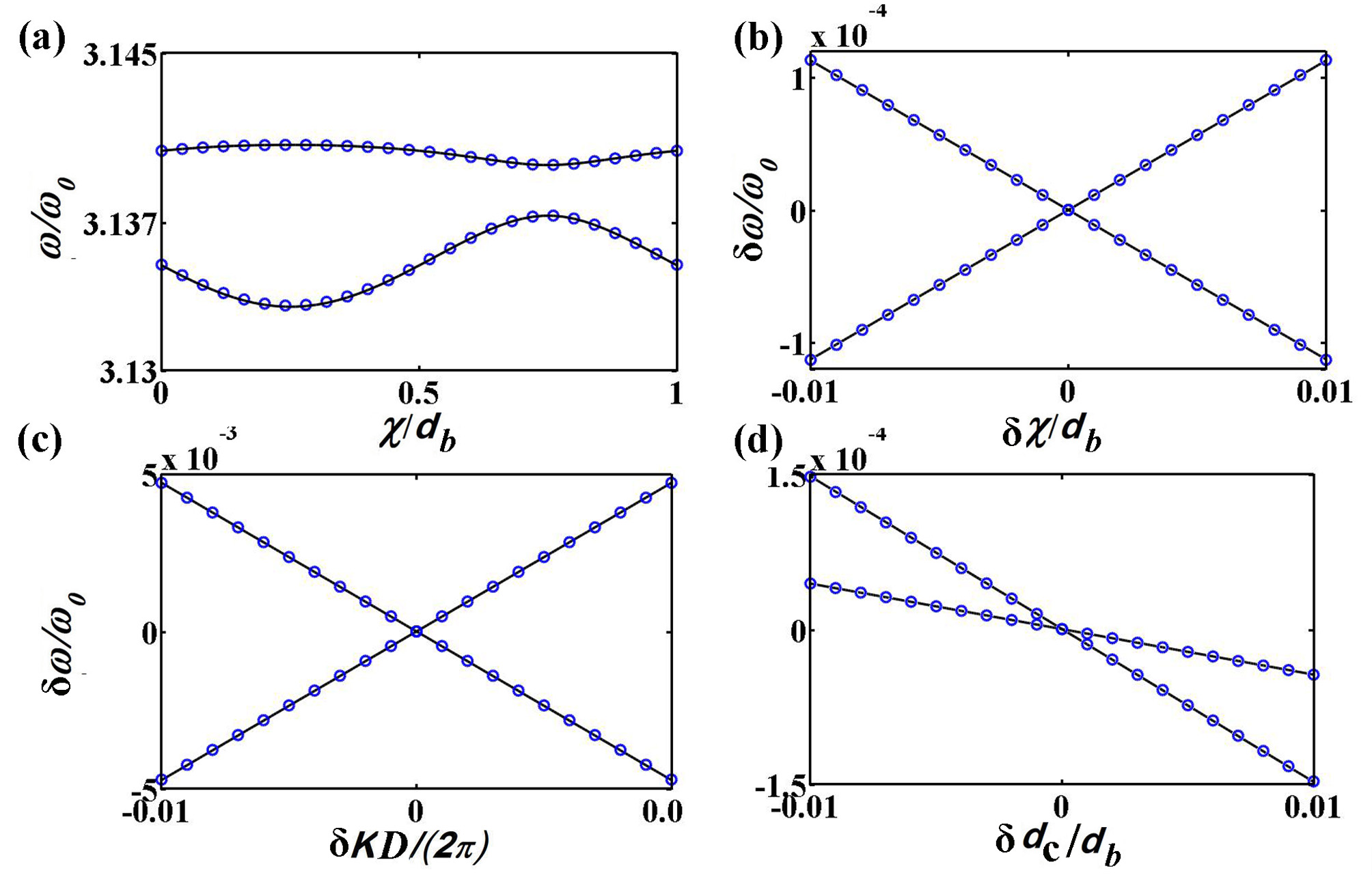}
\caption{\label{fig_6} Panels (a)-(d) demonstrate the comparison between the effective Hamiltonian(solid line) in Eq. \ref{eq:5}, \ref{eq:12} and the numerical result. Panel (a) is the dispersion of the second gap along $\chi$ ,which is shown in Fig. \ref{fig_4} (c). Panels (c)-(d) are the dispersion near the Weyl point in three directions, which is shown in Fig. \ref{fig_5} (d).}
\end{figure}
To summary, we propose a model with synthetic dimensions and very rich topological properties are investigated. We find that almost all non-trivial topological properties of 2D synthetic system are determined by the 1D-SSIS, such as, the topology triviality of 2D synthetic band-gap could be judged by the parity-switching types of gap-edge states at two 1D-SSIS, the Chern number of upper (or lower) band of the second  gap can be calculated by the Zak phases of 1D-SSIS. The deep physical picture is that there are two singular points in the upper or lower bands at two 1D-SSIS, which dominate not only topology (such as Zak phases and parity-switching type of gap-edge states) of 1D-SSIS, but also the topology of 2D synthetic system and higher-dimension systems.
Two types of topological transition are studied for such simple model. In the first type, we find that two bands touch each other as the Weyl point, which acts as the instantaneous jumping channel from one band to the other for one singularity of 1D-SSIS. In the second type, both singularities from two bands are merged into a ``nodal line" and emerge when the condition of nodal line is passed. A theory is constructed  which could quantitatively describe the topology for the systems with different dimensions.Further questions after this work are still waiting to answer, such as "Why a low-dimension system with certain symmetry can dominates the topological properties of the systems with higher-dimension?", "Is this a general phenomenon for other models? ", and "For Weyl-point-type and nodal-line-type topological transitions, the evolving processes of singularities revealed in this paper are universal?". Finally, we note that our model could be realized in laboratory and experimentally checked. With both layer B and layer C as liquid material contained in half-open and width-tunable boxes, the layer A which is a solid slab could be moved through the whole cell easily.
\\

\section*{ACKNOWLEDGMENTS}
This work was supported by a grant from the National Nature Science Foundation of China(11334015), a grant from the National Key Research Program of China (2016YFA0301103,2018YFA0306201) and National High Technology Research and Development Program of China(863Progtam)(17-H863-04-ZT-001-035-01).

\bibliography{manuref.bib}

\end{document}